\documentclass[english,aps,pra,10pt,tightenlines,fleqn,twocolumn,groupedaddress,floatfix]{revtex4-1}
\pdfoutput=1
\usepackage[utf8]{inputenc}
\usepackage[T1]{fontenc}
\usepackage{amsmath}
\usepackage{amssymb}
\usepackage{amsfonts}
\usepackage{bm}
\usepackage{bbm}
\usepackage{epsfig}
\usepackage{grffile}
\usepackage{graphics}
\usepackage{scalefnt}

\usepackage{wasysym}

\usepackage[usenames,dvipsnames]{color}
\definecolor{dblue}{rgb}{0,0.1,.6}

\usepackage[colorlinks=true,citecolor=dblue,linkcolor=dblue,urlcolor=dblue]{hyperref}
\usepackage[all]{hypcap}

\newcommand{\id}{\mathbbm{1}}
\newcommand{\bra}{\langle}
\newcommand{\ket}{\rangle}
\newcommand{\bbra}{\langle\!\langle}
\newcommand{\kket}{\rangle\!\rangle}
\newcommand{\Tr}{\operatorname{Tr}}
\renewcommand{\vec}[1]{{\boldsymbol{#1}}}
\newcommand{\ud}{\mathrm{d}}

\newcommand{\hH}{\hat{H}}
\newcommand{\hL}{\hat{L}}
\newcommand{\hM}{\hat{M}}
\newcommand{\dm}{{\hat{\rho}}}
\newcommand{\ha}{\hat{a}}
\newcommand{\hw}{\hat{w}}
\newcommand{\hvw}{\hat{\vec{w}}}
\newcommand{\mri}{\mathrm{i}}
\newcommand{\va}{\vec{a}}
\newcommand{\vn}{\vec{n}}
\newcommand{\vk}{\vec{k}}
\renewcommand{\vr}{\vec{r}}
\newcommand{\vg}{\vec{g}}
\newcommand{\vgamma}{\vec{\gamma}}
\newcommand{\tgamma}{\tilde{\gamma}}
\newcommand{\tb}{\tilde{b}}
\newcommand{\tell}{\tilde{\vec{\ell}}}
\newcommand{\tx}{\tilde{x}}
\newcommand{\ty}{\tilde{y}}

\newcommand{\RR}{\mathbb{R}}
\newcommand{\CC}{\mathbb{C}}
\newcommand{\NN}{\mathbb{N}}
\newcommand{\ZZ}{\mathbb{Z}}
\renewcommand{\Re}{\operatorname{Re}}
\newcommand{\groupO} {\operatorname{O}}

\newcommand{\mc}[1]{\mathcal{#1}}
\newcommand{\pdag}{{\phantom{\dag}}}

\renewcommand{\L}{\mc{L}}
\newcommand{\D}{\mc{D}}
\newcommand{\N}{\mc{N}}
\newcommand{\Ns}{N}
\newcommand{\tr}{\mathrm{r}}
\newcommand{\ti}{\mathrm{i}}
\newcommand{\veps}{\varepsilon}

\usepackage{amsthm}
\newtheorem{proposition}{Proposition}

\renewcommand{\Bmatrix}[1]{\begin{bmatrix}#1\end{bmatrix}}
\newcommand  {\Pmatrix}[1]{\begin{pmatrix}#1\end{pmatrix}}
\newcommand  {\Psmatrix}[1]{\left(\begin{smallmatrix}#1\end{smallmatrix}\right)}

\newcommand{\duke} {Department of Physics, Duke University, Durham, North Carolina 27708, USA}
\newcommand{\Title} {Criticality and Phase Classification for Quadratic Open Quantum Many-Body Systems}
\newcommand{\Authors}
{
\author{Yikang Zhang}
\affiliation{\duke}
\author{Thomas Barthel}
\affiliation{\duke}
}
\newcommand{\Date} {August 1, 2022}

\begin{document}

\title{\Title}
\Authors
\date{\Date}

\begin{abstract}
We study the steady states of translation-invariant open quantum many-body systems governed by Lindblad master equations, where the Hamiltonian is quadratic in the ladder operators, and the Lindblad operators are either linear or quadratic and Hermitian. These systems are called quasifree and quadratic, respectively.
We find that steady states of one-dimensional systems with finite-range interactions necessarily have exponentially decaying Green's functions. For the quasifree case without quadratic Lindblad operators, we show that fermionic systems with finite-range interactions are noncritical for any number of spatial dimensions and provide bounds on the correlation lengths. Quasifree bosonic systems can be critical in $D>1$ dimensions. Last, we address the question of phase transitions in quadratic systems and find that, without symmetry constraints beyond invariance under single-particle basis and particle-hole transformations, all gapped Liouvillians belong to the same phase.
\end{abstract}

\maketitle

\section{Introduction}
For closed systems, criticality and quantum phase transitions have been studied extensively \cite{Sachdev2011,Vojta2003-66,Sondhi1997-69,Zeng2019}. Particularly, for one-dimensional systems, we have obtained a thorough classification of gapped states using the tensor-network ansatz \cite{Chen2011-83,Schuch2011-84b,Chen2011-84,Cirac2021-65}.

In practice, most quantum systems are not perfectly isolated from their environment. In addition to posing challenges for the realization of quantum technology, driving and dissipation in open systems could be designed to stabilize (novel) phases of matter or particular entangled states \cite{Kraus2008-78,Diehl2008-4,Verstraete2009-5}, e.g., to facilitate measurement-based quantum computation \cite{Raussendorf2001-86,Walther2005-434}, quantum phase estimation \cite{Abrams1999-83,AspuruGuzik2005-309}, and quantum simulation \cite{Feynman1982-21,Bohnet2016-352,Gross2017-357,Bernien2017-551,Browaeys2020-16}.
For Markovian systems, the density matrix $\dm$ evolves according to a Lindblad master equation \cite{Lindblad1976-48,Gorini1976-17,Breuer2007,Rivas2012,Wolf2008-279}
\begin{equation*}
	\partial_t \dm=\L\dm=-i[\hH,\dm]+\sum_{\alpha}\left(\hL_\alpha\dm \hL_\alpha^\dag-\frac{1}{2}\{\hL_\alpha^\dag \hL_\alpha,\dm\}\right).
\end{equation*}
In addition to the Hamiltonian part $-i[\hH,\dm]$, the Liouvillian superoperator $\L$ captures decoherence processes with environment couplings described by the Lindblad operators $\hL_\alpha$.

In this work, we elucidate the occurrence of criticality and phase transitions in the steady states of open quasifree and quadratic systems of fermions and bosons. Quasifree open systems are characterized by Hamiltonians that are bilinear and Lindblad operators that are linear in ladder operators. Quadratic open systems may have additional bilinear self-adjoint Lindblad operators \cite{Barthel2021_12,Horstmann2013-87}. A system is called \emph{critical} if it has a unique steady state with algebraically decaying correlations. We establish that quadratic one-dimensional (1D) systems with finite-range interactions and unique steady states necessarily have exponentially decaying Green's functions (Sec.~\ref{sec:quadratic1d}). Next, we address quasifree systems with finite-range interactions. Quasifree fermionic systems are noncritical for any number of spatial dimensions (Sec.~\ref{sec:quasifree}) \footnote{Ref.~\cite{Eisert2010_12} contains an incomplete argument for 1D quasifree fermionic systems.}.
Conversely, one can construct critical quasifree bosonic systems for $D\geq 2$ dimensions. Gapped quasifree systems are always noncritical. Of course, the existence of critical steady states does not necessarily imply phase transitions. In fact we show in Sec.~\ref{sec:phases} that, without symmetry constraints beyond invariance under single-particle basis and particle-hole transformations, all gapped Liouvillians of quadratic open systems belong to the same phase.

Experimentally, systems of trapped ions \cite{Cirac1995-74,Blatt2008-453}, Rydberg atoms \cite{Jaksch2000-85,Lukin2001-87}, ultracold atoms in optical lattices or tweezers  \cite{Bloch2007,Norcia2018-8,Cooper2018-8}, and superconducting circuits \cite{Schoelkopf2008-451,Devoret2013-339} allow for the engineering of such dissipative systems \cite{Weimer2010-6,Barreiro2011-470,Mueller2012-61,Leghtas2013-88,Carr2013-111b,Rao2013-111}.
In circuit QED systems \cite{Hartmann2006-2,Angelakis2007-76,Hartmann2016-18,Fitzpatrick2017-7}, linear Lindblad operators arise naturally from photon loss and pump process, while the coupling of cavities can lead to bilinear Lindblad operators \cite{Marcos2012-14,Tomadin2012-86}.

\section{Setup and covariance matrix}\label{sec:setup}
Consider a system of identical bosons or fermions with ladder operators $\ha_j$ and $\ha^\dag_j$ for modes $j=1,\dotsc,\Ns$. We employ Majorana operators $\hw_{j+}:=(\ha_j+\ha_j^\dag)/\sqrt{2}$ and  $\hw_{j-}:=i(\ha_j-\ha_j^\dag)/\sqrt{2}$ which obey the (anti-)commutation relations
\begin{align*}
	\{\hw_{i\mu},\hw_{j\nu}\}&=\delta_{i,j}\delta_{\mu\nu} \quad &&\text{for fermions, and}\\
	[\hw_{i\mu},\hw_{j\nu} ]&=-\mri\mu\,\delta_{i,j}\delta_{\mu,-\nu} \quad &&\text{for bosons}.
\end{align*}
We address Markovian systems with quadratic Hamiltonians
$\hH=\sum_{i\mu,j\nu}\hw_{i\mu} H_{i\mu,j\nu}\hw_{j\nu}$.
Quasifree systems only have linear Lindblad operators $\hL_s=\sum_{j\nu}L_{s,j\nu}\hw_{j\nu}$.
Quadratic systems may feature additional bilinear self-adjoint Lindblad operators
$\hM_u=\hM_u^\dag=\sum_{i\mu,j\nu}\hw_{i\mu} (M_u)_{i\mu,j\nu}\hw_{j\nu}$.
The $2\Ns\times 2\Ns$ covariance matrix
\begin{equation}\label{eq:Cov}
	\Gamma_{i\mu,j\nu} :=
	\begin{cases}
	\frac{\mri}{2}\bra\hw_{i\mu}\hw_{j\nu}-\hw_{j\nu}\hw_{i\mu}\ket &\text{for fermions},\\
	\frac{1}{2}\bra\hw_{i\mu}\hw_{j\nu}+\hw_{j\nu}\hw_{i\mu}\ket    &\text{for bosons}
	\end{cases}
\end{equation}
can be shown to evolve according to the equation of motion \cite{Horstmann2013-87,Barthel2021_12}
\begin{equation}\label{eq:EOM}
	\partial_t \Gamma = X\Gamma+\Gamma X^T + Y + \sum_u Z_u\Gamma Z_u^T,
\end{equation}
where the real $2\Ns\times 2\Ns$ matrices $X$, $Y$, and $Z_u$ depend on the coupling coefficients $H$, $L_s$, and $M_u$ as detailed in Appx.~\ref{sec:ti-quad-r}. The $Z_u$ term vanishes for quasifree systems.

For a translation-invariant system in $D$ dimensions, each mode $i$ is associated with a cell location $\vec{i}\in\ZZ^D$ and a crystal-basis index $c_i=1,\dotsc,b$, where $b$ is the number of bands. The covariance matrix elements and coupling coefficients are then functions of spatial distances such that
\begin{equation*}\hspace{-1ex}
	\Gamma_{i\mu,j\nu}=:\gamma_{c_i\mu,c_j\nu}(\vec{i}-\vec{j}),\quad
	X_{i\mu,j\nu}=:x_{c_i\mu,c_j\nu}(\vec{i}-\vec{j})
\end{equation*}
etc., and the equation of motion \eqref{eq:EOM} takes the form
\begin{multline}\label{eq:EOMti}\hspace{-3ex}
	\partial_t\gamma(\vr)
	= \sum_{\vn}\left[x(\vn)\gamma(\vr-\vn)+\gamma(\vr+\vn) x^T(\vn)\right] + y(\vr) \\\hspace{-2.5ex}
	  + \!\sum_{u,\vn,\vec{j},\vec{l}}\! z_u(\vr-\vn,\vec{j}-\vn)\gamma(\vr-\vec{l}) z_u^T(-\vn,\vec{l}-\vn),
\end{multline}
where $\gamma$, $x$, $y$, and $z_u$ are $2b\times 2b$ matrices depending on lattice translation vectors $\vr\in\ZZ^D$.

\section{Correlations in quadratic 1D systems}\label{sec:quadratic1d}
As a first result, let us establish the following.
\begin{proposition}\label{prop:1dQuad}
If a quadratic 1D system with translation-invariant finite-range couplings has a unique steady state, then its single-particle Green's function $\gamma(r)$ cannot follow a power-law decay with respect to the distance $|r|$.
\end{proposition}
For the steady-state covariance matrix $\gamma(r)$, the right-hand side of Eq.~\eqref{eq:EOMti} needs to be zero. For distances $r$ large enough such that the local $z_u$ and $y$ terms vanish, $\gamma(r)$ obeys a matrix difference equation of the form
\begin{equation}\label{eq:1dMatrixDiff}
	C_0 \vgamma(r) + C_1 \vgamma(r+1)+\dotsb +C_R \vgamma(r+R)=\vec{0}.
\end{equation}
Here $\vgamma(r)$ is the vectorization of $\gamma(r)$, the $4b^2\times 4b^2$ matrices $C_m$ are determined by the coupling matrices $x(n)$, and $R$ denotes the interaction range; cf.~Appx.~\ref{sec:1dQuad}.

In the simplest scenario, $C_R$ is invertible such that we can solve Eq.~\eqref{eq:1dMatrixDiff} for $\vgamma(r+R)$ and 
\begin{equation}\label{eq:1dTransferMatrix}
	\vec{g}_{r+1}
	=
	\Bmatrix{
	A_{R-1}  & A_{R-2}  & \cdots & A_{1}  & A_{0} \\
	\id      & 0        & \cdots &  0     & 0 \\
	0        & \id      & \cdots &  0     & 0 \\
	\vdots   & \vdots   & \ddots & \vdots & \vdots \\
	0        & 0        & \cdots & \id    & 0
	}
	\vec{g}_{r}
\end{equation}
with $\vec{g}^T_r:=(\vgamma^T(r+R-1),\dotsc,\vgamma^T(r))$ and $A_m:=-C_R^{-1}C_m$. The spectrum of the $4b^2 R\times 4b^2R$ transfer matrix in Eq.~\eqref{eq:1dTransferMatrix} characterizes the spatial decay of $\gamma(r)$. As the spectrum is discrete, all elements of $\gamma(r)$ must decay exponentially, converge to a constant, or oscillate with constant amplitude. An algebraic decay that characterizes critical systems is not possible. The transfer matrix may have eigenvalues $\beta$ with $|\beta|>1$. These are, however, irrelevant as physical systems cannot feature indefinitely growing $\gamma(r)$. For fermions, this is also prohibited by the constraint that all covariance matrix elements lie in the interval $[-1/2,1/2]$ \cite{Barthel2021_12}.
Appendix~\ref{sec:1dQuad} gives a more general proof based on generating functions, which does not require invertibility of $C_R$.

\section{Criticality in quasifree systems}\label{sec:quasifree}
Stronger results hold for the systems that have no quadratic Lindblad operators and, hence, no $Z_u$ term in Eq.~\eqref{eq:EOM}. Let us first consider \emph{gapped} systems, where the Liouvillian $\L$ has a single zero eigenvalue and the other eigenvalues $\lambda$ have a nonzero \emph{dissipative gap} $\Delta:=-\max_{\lambda\neq 0}\Re\lambda>0$.
\begin{proposition}\label{prop:gapped}
Gapped quasifree systems with translation-invariant finite-range couplings are never critical.
\end{proposition}
Note that, using quasilocality \cite{Barthel2012-108b}, this proposition can be generalized to interacting systems. But quasifree systems allow for a more direct proof that provides bounds on correlation lengths to be reused for Prop.~\ref{prop:quasiFreeF}:

Because of translation invariance, we can transform to a momentum-space representation with quasimomenta $k_a=\frac{2\pi}{L},\frac{4\pi}{L},\dotsc,2\pi$ for $a=1,\dotsc,D$. With
\begin{equation}\label{eq:qftiFourier}
	\tgamma(\vk):=\sum_\vr e^{-\mri \vk\cdot\vr}\gamma(\vr),\quad
	\tx(\vk):=\sum_\vr e^{-\mri \vk\cdot\vr}x(\vr)
\end{equation}
and an analogous definition of $\ty$, according to Eq.~\eqref{eq:EOMti}, the steady state obeys the continuous Lyapunov equation
\begin{equation}\label{eq:qftiSteadyState}
	\tx(\vk)\tgamma(\vk)+\tgamma(\vk)\tx^T(-\vk)=-\ty(\vk).
\end{equation}
For a quasifree system to be gapped, all eigenvalues of $X$ in Eq.~\eqref{eq:EOM} or, equivalently, all eigenvalues of $\tx(\vk)$ $\forall \vk$ in Eq.~\eqref{eq:qftiSteadyState} need to have negative real parts \cite{Barthel2021_12}. But this means that we can solve Eq.~\eqref{eq:qftiSteadyState} for $\tgamma(\vk)$ by inverting the matrix $\tx(\vk)\otimes \id + \id\otimes \tx(-\vk)$. Due to the finite interaction range, $\tx(\vk)$ and $\ty(\vk)$ are polynomials in variables $z_a:=e^{\mri k_a}\in \CC$ and $1/z_a$. Hence, $\tgamma(\vk)$ is a rational function of the $z_a$ which, according to the invertibility of $\tx(\vk)$, has no poles on the manifold $|z_a|=1$ which corresponds to real momenta $k_a\in(0,2\pi]$ in the Brillouin zone. For concreteness, let us discuss $D=2$ dimensions; the generalization to $D\neq 2$ is trivial. The established property of $\tgamma(\vk)=:\tgamma(z_1,z_2)$ allows us to determine $\gamma(r_1,r_2)$ using Cauchy's residue theorem from complex analysis:

In the thermodynamic limit, the inverse of Eq.~\eqref{eq:qftiFourier} is
\begin{equation}
	\gamma(r_1,r_2)= -\!\!\!\!\!\!\!\oiint_{|z_1|=|z_2|=1} \!\!\frac{\ud^2 z}{(2\pi)^2}\,\, z_1^{r_1-1}z_2^{r_2-1}\tgamma(z_1,z_2).
\end{equation}
For fixed $z_2$, let $\varrho(z_2):= \mri\sum_m \operatorname{Res}\big(\tgamma(\zeta_m(z_2),z_2)\big)$ denote the sum over the residues of $\tgamma$ at pole locations $z_1=\zeta_m(z_2)$ inside the unit circle $|z_1|=1$ \footnote{This statement and the following considerations actually concern matrix elements of $\gamma$ and $\tgamma$, but we drop the matrix indices for brevity of notation.}. With $|\zeta|:=\max_{m,|z_2|=1}|\zeta_m(z_2)|<1$, it follows that 
\begin{equation}
	|\gamma(r_1,r_2)|\leq |\zeta|^{r_1-1} \oint_{|z_2|=1} \frac{\ud z_2}{2\pi}\,\left|z_2^{r_2-1} \varrho(z_2)\right|.
\end{equation}
As the contour integral is independent of $r_1$, this bound establishes an exponential decay of $\gamma(r_1,r_2)$ with correlation length
\begin{equation}\label{eq:qftiCorrLength}
	\xi_1\leq -1/\ln|\zeta| = - 1\,/\max_{m,|z_2|=1} \ln |\zeta_m(z_2)|
\end{equation}
in the positive $r_1$ direction. An exponential bound for negative $r_1$ is obtained by using $z_1:=e^{-\mri k_1}$ instead of $e^{\mri k_1}$, and the same arguments apply to $r_2$ or further dimensions.

The steady states of quasifree systems are Gaussian \cite{Barthel2021_12}. Hence, according to Wick's theorem \cite{Wick1950-80,Negele1988}, the steady state is fully characterized by $\gamma(\vr)$, and the exponential decay of $\gamma(\vr)$ implies the exponential decay of all connected real-space correlation functions. This concludes the proof of Prop.~\ref{prop:gapped}.
Let us now drop the constraint of a nonzero dissipative gap.

\begin{proposition}\label{prop:quasiFreeF}
Quasifree fermionic systems with translation-invariant finite-range couplings are never critical.
\end{proposition}
For a unique steady state, the momentum-space covariance matrix $\tgamma(\vk)$ solving Eq.~\eqref{eq:qftiSteadyState} is again a rational function. Furthermore, it cannot have poles at real $\vk$ for any short-range fermionic system \cite{Note1}:
The covariance matrix $\Gamma$ in Eq.~\eqref{eq:Cov} is real and anti-symmetric. Hence, there exists an orthogonal transformation $O\in\groupO(2\Ns)$ such that $\Gamma':=O\Gamma O^T=\Psmatrix{&\chi\\-\chi&}$, where the elements $\chi_{i}$ of the $\Ns\times\Ns$ diagonal matrix $\chi$ correspond to the imaginary eigenvalue pairs $\pm \mri\chi_{i}$. The transformation defines an alternative set of Majorana operators $\hw'_{i\mu}:=\sum_{j\nu} O_{i\mu,j\nu}\hw_{j\nu}$ with covariance matrix $\Gamma'$ such that $\chi_{i}=\mri\bra \hw'_{i+}\hw'_{i-}\ket$. As each fermionic occupation number operator $\ha_j^\dag\ha^\pdag_j$ has eigenvalues $0$ and $1$, the operators $\mri\hw_{j+}\hw_{j-}={1}/{2}-\ha_j^\dag\ha^\pdag_j$ and the operators $\mri\hw'_{i+}\hw'_{i-}$ have eigenvalues $\pm 1/2$. Thus, all $\chi_i$ are in the interval $[-{1}/{2},{1}/{2}]$, and all covariance matrix elements obey $|\Gamma_{i,j}|\leq \|O^T\Gamma' O\|=\|\chi\oplus(-\chi)\|\leq \frac{1}{2}$. The Fourier transform \eqref{eq:qftiFourier} to momentum space just adds another unitary transformation. Hence, the elements of $\tgamma(\vk)$ have modulus $\leq 1/2$, i.e., singularities can only occur at complex momenta $k_a$. Their imaginary parts provide bounds on correlation lengths as in Eq.~\eqref{eq:qftiCorrLength}, and the system is not critical.

Proposition~\ref{prop:quasiFreeF} is in stark contrast to closed fermionic systems, where tight-binding models have, for example, critical Fermi-sea ground states.
The situation for open bosonic systems is different. Note that bosonic open systems can be unstable in the sense that the Liouvillian can have eigenvalues with positive real parts that lead to unlimited absorption of energy and particles. In quasifree systems, however, the existence of a steady state implies stability \cite{Barthel2021_12}. So, stability is implied in the following.
\begin{proposition}\label{prop:quasiFreeB}
Quasifree bosonic systems with translation-invariant finite-range couplings can be critical in $D\geq 2$ dimensions. 1D systems cannot be critical.
\end{proposition}
The statement on 1D systems follows immediately from Prop.~\ref{prop:1dQuad} and Wick's theorem. Furthermore, one can construct quasifree bosonic models that are critical for $D\geq 2$ dimensions. Specifically, consider a purely dissipative model with one Lindblad operator
  $\hL^{(1)}_\vec{j}:=\sqrt{2D\eta}(\hw_{\vec{j}+}-\mri \hw_{\vec{j}-})=\sqrt{2D\eta}\,\ha_\vec{j}$
for every site $\vec{j}\in\ZZ^D$ of the $D$-dimensional square lattice as well as four Lindblad operators
  $\hL^{(2\pm)}_{\vec{j},a}:=\hw_{\vec{j}+}+\mri \hw_{(\vec{j}\pm\vec{e}_a)-}$ and
  $\hL^{(3\pm)}_{\vec{j},a}:=\hw_{\vec{j}+}\pm \hw_{(\vec{j}\pm\vec{e}_a)-}$
for every edge, where $\vec{e}_a$ are the unit vectors for directions $a=1,\dotsc,D$.
One finds that $\tx(\vk)=2D\,(c_\vk-\eta) \id_2$, where $c_\vk:=\sum_a\cos k_a/D$; cf.~Appx.~\ref{sec:criticalBosons}. The largest $X$ eigenvalue real part determines the dissipative gap $\Delta$ \cite{Barthel2021_12}. Here, $\tx(\vk)$ has the doubly degenerate eigenvalue $\xi(\vk)=2D\,(c_\vk-\eta)$ and, hence, $\Delta=-\max_\vk\Re\xi(\vk)=2D\,(\eta-1)$. So the model is stable for loss rates $\eta\geq 1$ and the gap closes for $\eta=1$ at momentum $\vk=\vec{0}$. Solving the Lyapunov equation \eqref{eq:qftiSteadyState} yields the covariance matrix $\tgamma(\vk)$ with the diagonal and off-diagonal elements 
\begin{equation}\label{eq:criticalBosons-gk}
	\tgamma_{\pm,\pm}(\vk)=\frac{\eta+2}{2(\eta-c_\vk)}\,\ \text{and}\,\
	\tgamma_{\pm,\mp}(\vk)=\frac{\pm\mri s_\vk}{2(\eta-c_\vk)},
\end{equation}
where $s_\vk:=\sum_a\sin k_a/D$. With a Fourier transform to $\gamma(\vr)$, one can assess criticality. For $D=1$ dimensions, the Fourier integral can be evaluated exactly using the residue theorem. In agreement with Props.~\ref{prop:1dQuad} and \ref{prop:gapped}, we find an exponential decay of correlations if $\eta>1$. The correlation length diverges for $\eta\to 1$, but there is no power-law decay. For dimensions $D\geq 2$, one can expand $\tgamma(\vk)$ in a multipole series over hyperspherical harmonics \cite{Avery2017} to reduce the Fourier transformation to a radial integral, which takes the form of a Hankel transform. The leading contributions to $\tgamma_{\pm,\pm}$ are isotropic while those to $\tgamma_{\pm,\mp}$ are anti-symmetric with respect to reflection. For $D=2$ dimensions, the diagonal correlations $\gamma_{\pm,\pm}(\vr)$ decay logarithmically in $|\vr|$ and the off-diagonal $\gamma_{\pm,\mp}(\vr)$ decay as $1/|\vr|$. For $D=3$, they decay as $1/|\vr|$ and $1/|\vr|^2$, respectively. A detailed discussion is given in Appx.~\ref{sec:criticalBosons}.

\section{Phase classification for quadratic systems}\label{sec:phases}
Like quantum phase transitions in closed systems \cite{Sachdev2011,Vojta2003-66,Sondhi1997-69}, driven-dissipative phase transitions are characterized by a nonanalytic dependence of steady-state expectation values on system parameters. This requires a nonanalytic change in the steady-state density matrix and, hence, a level crossing \cite{Kato1995}. So, the dissipative gap $\Delta$ needs to close at the transition point \cite{Kessler2012-86,Minganti2018-98}. As seen so far, there are some restrictions on criticality in quadratic open systems, but the gap can of course close. As another fundamental result, we will see why, here, closing the gap does generally not lead to phase transitions.
\begin{proposition}\label{prop:phaseUnity}
For quadratic systems without symmetry constraints beyond invariance under single-particle basis transformations and fermionic particle-hole symmetry, all gapped systems belong to the same phase. For any pair of gapped systems $\L_1$ and $\L_2$, one can construct a continuous path of gapped Liouvillians that links the two.
\end{proposition}
In particular, we claim that for any quadratic Liouvillian $\L$ with gap $\Delta$, the auxiliary Liouvillian
\begin{equation}\label{eq:auxL-gap}
	\L+\kappa \D\quad\text{has a gap}\quad \Delta'\geq\Delta+\kappa.
\end{equation}
For fermionic systems, the added dissipator $\D$ comprises two linear Lindblad operators $\hL_{i\pm}=\hw_{i\pm}$ for every mode $i$ \footnote{A \emph{dissipator} \unexpanded{$\D$} comprising Lindblad operators \unexpanded{$\hL_1,\dotsc,\hL_n$} acts as \unexpanded{$\D\dm=\sum_{\alpha=1}^n\big(\hL_\alpha\dm \hL_\alpha^\dag-\frac{1}{2}\{\hL_\alpha^\dag \hL_\alpha,\dm\}\big)$}.}. For bosons, $\D$ comprises one operator $\hL_i=\hw_{i+}-\mri\hw_{i-}=\sqrt{2}\,\ha_i$ per mode. With this choice and any $\kappa_0>0$, the gap stays nonzero, e.g., along the path $(1-g)\L_1+g\L_2+\kappa\D$, where the parameters are tuned as $(g,\kappa):(0,0)\to(0,\kappa_0)\to(1,\kappa_0)\to(1,0)$ to connect $\L_1$ to $\L_2$.
Note that this proposition does not require short-range interactions.

Statement \eqref{eq:auxL-gap} can be proven by employing the third-quantization formalism \cite{Prosen2008-10,Prosen2010-07,Prosen2010-43,Barthel2021_12} as detailed in the companion paper \cite{Barthel2021_12}: (a)
There exist ladder superoperators $a_{j\nu}$ and $a'_{j\nu}$ that obey canonical (anti-)commutation relations and form a basis for the superoperator algebra. (b)
One can then construct a biorthogonal operator basis
\begin{equation}\label{eq:opBasis}
	\bbra\vn|\quad\text{and}\quad|\vn\kket\quad\text{with}\quad \bbra\vn|\vn'\kket=\delta_{\vn,\vn'},
\end{equation}
occupation numbers $\vn^T=(n_{1+},\dotsc,n_{\Ns+},\dotsc,n_{\Ns-})$, and $a'_{j\nu}a_{j\nu}|\vn\kket=n_{j\nu}|\vn\kket$. The Dirac notation with super-bras $\bbra\hat{A}|$ and super-kets $|\hat{B}\kket$, where $\hat{A}$ and $\hat{B}$ are operators on the Hilbert space, is based on the Hilbert-Schmidt inner product $\bbra \hat{A}|\hat{B}\kket\equiv \Tr(\hat{A}^\dag\hat{B})$. (c)
The ladder superoperators can be chosen such that the matrix representation $\bbra\vn|\L|\vn'\kket$ of $\L$ assumes a block-triangular form when ordering the basis \eqref{eq:opBasis} according to increasing eigenvalues $N_a\in\NN$ of the number superoperator $\N_a:=\sum_{j\nu} a'_{j\nu}a_{j\nu}$. The spectra of the blocks $\L|_{N_a}$ on the diagonal determine the full Liouvillian spectrum \cite{Barthel2020_12}. The only terms due to $\D$ that affect the blocks $\L|_{N_a}$ are \cite{Barthel2021_12}
\begin{equation}\label{eq:auxL-affected}\textstyle
	-\va'^{T}\frac{B+B^*}{2}\va,\ \
	-\va^{T}\frac{B+B^*}{2}\va',\ \
	\va'^{T}U^\dag\tau \frac{B-B^*}{2}U\va
\end{equation}
for fermions with even $N_a$, fermions with odd $N_a$, and bosons, respectively. In Eq.~\eqref{eq:auxL-affected}, $\va^T=(a_{1+},\dotsc,a_{\Ns-})$ and $\va'^T=(a'_{1+},\dotsc,a'_{\Ns-})$ are vectors containing all ladder superoperators, $U$ is a unitary matrix, $\tau=\Psmatrix{0&-\mri\id_{\Ns}\\\mri\id_{\Ns}&0}$, and $B_{i\mu,j\nu}=\sum_s L_{s,i\mu}L^*_{s,j\nu}$ is a positive-semidefinite matrix, characterized by the expansion coefficients of the linear Lindblad operators $\hL_s=\sum_{j\nu}L_{s,j\nu}\hw_{j\nu}$.

For fermions, the Lindblad operators of dissipator $\D$ have coefficients $L_{i\pm,j\nu}=\delta_{i,j}\delta_{\pm,\nu}$ and, hence, $B=\id_{2\Ns}$ such that the first two terms in Eq.~\eqref{eq:auxL-affected} are simply $-\N_a$ and $\N_a-2\Ns$, respectively. This implies that the spectrum of block $\L|_{N_a}$ is shifted by $-N_a\kappa$ and $(N_a-2\Ns)\kappa$ for even and odd $N_a$, respectively. As the $N_a=0$ block that contains the steady-state eigenvalue zero is one-dimensional, the spectral shifts due to $\kappa\D$ necessarily increase the gap to $\Delta'\geq\Delta+\kappa$. For bosons, we have $L_{i,j+}=\delta_{i,j}$ and $L_{i,j-}=-\mri\delta_{i,j}$. Hence, $B=\Psmatrix{\id_{\Ns}&\mri\id_{\Ns}\\-\mri\id_{\Ns}&\id_{\Ns}}$ and $\tau (B-B^*)/{2}=-\id_{2\Ns}$ such that the third term in Eq.~\eqref{eq:auxL-affected} reads $-\N_a$. Thus, also in the bosonic case, the gap increases at least by $\kappa$. For quasifree fermionic and bosonic systems, the gap increases exactly by $\kappa$, i.e., $\Delta'=\Delta+\kappa$.
Dissipator $\D$ is invariant under single-particle basis transformations $\ha_j\leftrightarrow \sum_i U_{j,i}\ha_i$ and also under particle-hole transformations $\ha_j\leftrightarrow \ha^\dag_j$ for fermions. This completes the proof of Prop.~\ref{prop:phaseUnity}.

\section{Example}\label{subsec_eg}
\begin{figure}[t]
	\centering
	\includegraphics[width=\columnwidth]{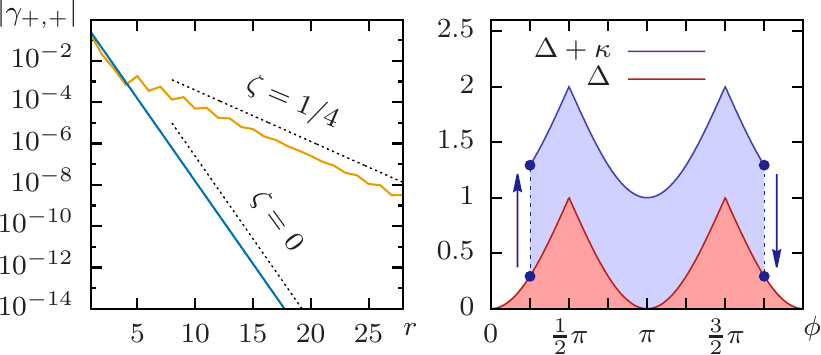}
	\caption{\label{fig:XYchain}The open fermionic model \eqref{eq:XYchain} with $\eta=1$ and $\mu=0$. Left: Both in the quasifree case ($\zeta=0$) and quadratic case ($\zeta=1/4$) with $\alpha=1/5$ and $\phi=2\pi/5$, correlations decay exponentially, where the asymptotic form $\sim \beta^r$ (dashed lines) is determined by an eigenvalue $\beta$ of the transfer matrix in Eq.~\eqref{eq:1dTransferMatrix}. Right: The dissipative gap $\Delta$ for $\zeta=0$ and $\alpha=1/2$ vanishes at $\phi=0,\pi$. It can be increased using the additional dissipator $\kappa\D$ from Eq.~\eqref{eq:auxL-gap}.}
\end{figure}
To illustrate some of the above results, consider the quadratic fermionic 1D model with Hamiltonian
\begin{equation}\label{eq:XYchain}
	\hH=\sum_{j}\big(\ha_j^\dag\ha_{j+1}^\pdag+\alpha\ha_j^\dag\ha_{j+1}^\dag +H.c.\big)
	-\mu\sum_j\ha_j^\dag\ha_j^\pdag,
\end{equation}
corresponding to a spin-1/2 XY chain, and Lindblad operators
$\hL_j=\sqrt{\eta}\,\left(\hw_{j+}+e^{\mri \phi}\hw_{(j+1)+}\right)$ as well as
$\hM_j=\sqrt{\zeta}\,\big(2\ha_j^\dag\ha_j^\pdag-1\big)$.
In accordance with Prop.~\ref{prop:1dQuad}, $\gamma(r)$ is always found to decay as $\beta^r$ for an eigenvalue $\beta$ of the transfer matrix in Eq.~\eqref{eq:1dTransferMatrix}.
Proposition~\ref{prop:quasiFreeF} implies that the quasifree model ($\zeta=0$), considered in Ref.~\cite{Eisert2010_12}, is never critical, and $\tx(k)$ determines the full many-body spectrum \cite{Barthel2021_12}. In particular, if the Hamiltonian is gapped and $\eta>0$, the dissipative gap $\Delta$ closes only at $\phi=0$ and $\pi$. The correlation length diverges at those points ($\beta\to 1$) but, at the same time, $\gamma(r)\to 0$ for all $r$. Furthermore, employing the additional dissipator $\kappa\D$ from Eq.~\eqref{eq:auxL-gap}, any two gapped points can always be connected by a path of gapped Liouvillians as explained by Prop.~\ref{prop:phaseUnity} and illustrated in Fig.~\ref{fig:XYchain} for the points $\phi=\pi/4$ and $\phi=9\pi/4$. So, the system is neither critical at $\phi=0$ or $\pi$, nor does it undergo phase transitions.
Details are presented in Appx.~\ref{sec:XYchain}.

\section{Discussion}\label{sec_discuss}
We have found fundamental prerequisites for criticality and phase transitions in driven-dissipative many-body systems that are in stark contrast to properties of closed systems. For any number of spatial dimensions, there exist fermionic and bosonic closed systems with phase transitions and critical ground states, i.e., states featuring an algebraic decay of spatial correlations, even if the systems are quasifree. In contrast, steady states of open 1D quasifree systems as well as higher-dimensional quasifree fermionic systems are never critical. For quadratic systems, we found that, while the dissipative gap may close and the system might even be critical for certain points in parameter space, all steady states basically belong to the same phase. The only way for realizing phase transitions in such systems is to impose symmetries on the considered Liouvillians that go beyond invariance under single-particle basis transformations (e.g., lattice symmetries) and fermionic particle-hole transformations or combinations thereof. A notable example are topological transitions in quasifree systems, occurring under the (strong) restriction that the Lindblad operators form a complete anti-commuting set \cite{Bardyn2013-15,Budich2015-91}. The observation that dissipative phase transitions are, in the above sense, more rare than phase transitions in closed systems adds to the idea that steady states are in certain scenarios related to thermal states of closed systems \cite{Mitra2006-97,Diehl2008-4,Rota2019-122,Young2020-10} such that continuous symmetries cannot be broken in $D\leq 2$ dimensions according to the Mermin-Wagner theorem \cite{Hohenberg1967-158,Mermin1966-17}. Interactions and more complex Lindblad operators can break the block-triangular Liouvillian structures \cite{Barthel2021_12,Barthel2020_12} that underlie our results on quasifree and quadratic systems and can cause true phase transitions \cite{Boite2013-110,Boite2014-90,Wilson2016-94,Savona2017-96,Rota2019-122,Verstraelen2020-2} as long as we are below an upper critical dimension where all systems become effectively quasifree.

\vspace{1em}
\begin{acknowledgments}\vspace{-1.2em}
We gratefully acknowledge discussions with Enej Ilievski, Israel Klich, Jianfeng Lu, and Toma\v{z} Prosen as well as
support through US Department of Energy grant DE-SC0019449.
\end{acknowledgments}

\appendix

\onecolumngrid

\section{Steady-state covariance matrix and translation invariance}\label{sec:ti}
\subsection{Quadratic systems in real-space representation}\label{sec:ti-quad-r}
As shown in Refs.~\cite{Barthel2021_12,Horstmann2013-87}, the covariance matrix $\Gamma\in\RR^{2\Ns\times 2\Ns}$ for a steady state of a quadratic open system obeys the linear equation
\begin{equation}\label{eq:SteadyState}\textstyle
	X\Gamma+\Gamma X^T + \sum_u Z_u\Gamma Z_u^T = -Y.
\end{equation}
The quadratic open systems are characterized by Hamiltonians $\hH$, which are quadratic in the Majorana operators $\hw_{j\nu}$, a set of linear Lindblad operators $\{\hL_s\}$, and a set of quadratic self-adjoint Lindblad operators $\{\hM_u\}$ such that
\begin{equation}\label{eq:H-L-M}
	\hH=\sum_{i\mu,j\nu}\hw_{i\mu} H_{i\mu,j\nu}\hw_{j\nu},\quad
	\hL_s=\sum_{j\nu}[\vec{L}_s]_{j\nu}\hw_{j\nu},\quad\text{and}\quad
	\hM_u=\hM_u^\dag=\sum_{i\mu,j\nu}\hw_{i\mu} [M_u]_{i\mu,j\nu}\hw_{j\nu}.
\end{equation}
Here, $H$ and $M_u$ are $2\Ns\times 2\Ns$ matrices, and $\vec{L}_s\in\CC^{2\Ns}$.
In addition, we define the positive semidefinite $2\Ns\times 2\Ns$ matrix $B$ with
\begin{equation}\label{eq:B}\textstyle
	B_{i\mu,j\nu}=\sum_s [\vec{L}_s^\pdag \vec{L}_s^\dag]_{i\mu,j\nu}= \sum_s [\vec{L}_s]_{i\mu}[\vec{L}_s^*]_{j\nu},\quad
	B_\tr:=\frac{B+B^*}{2}\quad \text{and}\quad
	B_\ti:=\frac{B-B^*}{2\mri}.
\end{equation}
For fermions, the real $2\Ns\times 2\Ns$ matrices in the steady-state equation \eqref{eq:SteadyState} are given by
\begin{equation}\label{eq:XYZ-F}\textstyle
	X=-2\mri H-B_\tr-2\sum_u M_u^2,\quad
	Y=B_\ti,\quad \text{and}\quad
	Z_u=2\mri M_u.
\end{equation}
For bosons, they are
\begin{equation}\label{eq:XYZ-B}\textstyle
	X=-2\mri\tau H+\mri\tau B_\ti-2\sum_u (\tau M_u)^2,\ \
	Y=\tau B_\tr\tau,\,\ \text{and}\,\
	Z_u=2\mri \tau M_u\,\ \text{with}\,\
	\tau=\Psmatrix{0&-\mri\id_{\Ns}\\\mri\id_{\Ns}&0}.
\end{equation}

In this work, we are concerned with translation invariant systems in $D$ dimensions. We can always choose a square lattice, where each fermionic or bosonic mode $i$ is associated with a lattice cell index $\vec{i}\in\{1,\dotsc,L\}^{\times D}$ and a crystal-basis index $c_i=1,\dotsc,b$. Here $b$ is the number of modes per unit cell (number of bands). With translation invariance, coupling coefficients and elements of the steady-state covariance matrix $\Gamma$ only depend on spatial distances in the sense that
\begin{equation}\label{eq:ti_H-G}
	H_{i\mu,j\nu}=:h_{c_i\mu,c_j\nu}(\vec{i}-\vec{j})\quad\text{and}\quad
	\Gamma_{i\mu,j\nu}=:\gamma_{c_i\mu,c_j\nu}(\vec{i}-\vec{j}).
\end{equation}
Both, $h(\vec{i}-\vec{j})$ and $\gamma(\vec{i}-\vec{j})$ are $2b\times 2b$ matrices.
For the dissipative terms to be translation invariant, we have Lindblad operators $\hL_{\vn,s}$ and $\hM_{\vn,u}$ that are centered at cell $\vn$ and related to those for cells $\vn'\neq\vn$ by lattice translations. Hence,
\begin{equation}\label{eq:ti_L-M}
	[\vec{L}_{\vn,s}]_{i\mu}=:[\vec{\ell}_s(\vec{i}-\vn)]_{c_i\mu} \quad\text{and}\quad
	[M_{\vn,u}]_{i\mu,j\nu}=:[m_u(\vec{i}-\vn,\vec{j}-\vn)]_{c_i\mu,c_j\nu}
\end{equation}
with $\vec{\ell}_s(\vec{i}-\vn)\in\CC^{2b}$ and $2b\times 2b$ matrices $m_u(\vec{i}-\vn,\vec{j}-\vn)$. For matrix \eqref{eq:B} follows
\begin{align}\nonumber
	B_{i\mu,j\nu} &\textstyle= \sum_{\vn,s} [\vec{L}_{\vn,s} \vec{L}_{\vn,s}^\dag]_{i\mu,j\nu}
	= \sum_{\vn,s} [\vec{\ell}_s(\vec{i}-\vn)]_{c_i\mu} [\vec{\ell}^*_s(\vec{j}-\vn)]_{c_j\nu}\\\label{eq:ti_B}
	&\textstyle= \sum_{\vn,s} [\vec{\ell}_s(\vec{i}-\vec{j}-\vn) \vec{\ell}^\dag_s(-\vn)]_{c_i\mu,c_j\nu}
	=: b_{c_i\mu,c_j\nu}(\vec{i}-\vec{j}).
\end{align}
With this, Eq.~\eqref{eq:SteadyState} assumes the manifestly translation invariant form
\begin{equation}\label{eq:ti_SteadyState}
	\sum_{\vn}\left[x(\vn)\gamma(\vr-\vn)+\gamma(\vr+\vn) x^T(\vn)\right]
	+ \!\sum_{u,\vn,\vec{j},\vec{l}}\! z_u(\vr-\vn,\vec{j}-\vn)\gamma(\vr-\vec{l}) z_u^T(-\vn,\vec{l}-\vn) =  -y(\vr),
\end{equation}
where $x$ and $y$ are related to $X$ and $Y$, respectively, like $h$ to $H$ [cf.~Eq.~\eqref{eq:ti_H-G}], and $z$ is related to $Z$ like $m$ to $M$ [cf.~Eq.~\eqref{eq:ti_L-M}].

As detailed in Ref.~\cite{Barthel2021_12}, all the $2N\times 2N$ matrices in Eq.~\eqref{eq:SteadyState} are real and have the following properties. For fermions, $\Gamma=-\Gamma^T$,  $H=H^\dag=-H^T$, and $M_u=M_u^\dag=-M_u^T$ such that the Hermitian component of $X$ is negative semidefinite, $Z_u=Z_u^T$, and $Y=-Y^T$. For bosons, $\Gamma=\Gamma^T$,  $H=H^\dag=H^T$, and $M_u=M_u^\dag=M_u^T$ such that $Z_u=-\tau Z_u^T\tau$ and $Y=Y^T$.
The $2b\times 2b$ matrices in the translation invariant representation \eqref{eq:ti_SteadyState} are then also real and have the following properties. For fermions, $\gamma(\vr)=-\gamma^T(-\vr)$, $h(\vr)=h^\dag(-\vr)=-h^T(-\vr)$, and $y(\vr)=-y^T(-\vr)$. For bosons, $\gamma(\vr)=\gamma^T(-\vr)$, $h(\vr)=h^\dag(-\vr)=h^T(-\vr)$, and $y(\vr)=y^T(-\vr)$.

\subsection{Quasifree systems in momentum-space representation}\label{sec:ti-quasifree-k}
For quasifree systems, where $Z_u=0$ as $M_u=0$, we will work in a momentum-space representation with quasimomenta $k_a=\frac{2\pi}{L},\frac{4\pi}{L},\dotsc,2\pi$ for spatial directions $a=1,\dotsc,D$. The Fourier transforms
\begin{equation}\label{eq:qfti_FourierAppx}\textstyle
	\tgamma(\vk):=\sum_\vr e^{-\mri \vk\cdot\vr}\gamma(\vr),\quad
	\tilde{h}(\vk):=\sum_\vr e^{-\mri \vk\cdot\vr}h(\vr),\quad\text{and}\quad
	\tell_s(\vk):=\sum_\vr e^{-\mri \vk\cdot\vr}\vec{\ell}_s(\vr)
\end{equation}
correspond to a block diagonalization of the matrices $\Gamma$, $H$, and $B$, where
\begin{equation}\label{eq:qfti_bk}\textstyle
	\tb(\vk):=\sum_\vr e^{-\mri \vk\cdot\vr}b(\vr)
	\stackrel{\eqref{eq:ti_B}}{=} \sum_s \tell_s(\vk) \tell^\dag_s(\vk),\quad
	\tb_\tr(\vk)=\frac{\tb(\vk)+\tb^*(-\vk)}{2}\ \ \text{and}\ \
	\tb_\ti(\vk)=\frac{\tb(\vk)-\tb^*(-\vk)}{2\mri}.
\end{equation}
Corresponding expressions apply for $x(\vr)\leftrightarrow \tx(\vk)$ and $y(\vr)\leftrightarrow\ty(\vk)$. Inserting into Eq.~\eqref{eq:ti_SteadyState}, one finds that the steady state obeys the \emph{continuous Lyapunov equation} \eqref{eq:qftiSteadyState}.
Lyapunov equations \cite{Parks1992-9,Simoncini2016-58} play an important roles in control theory and stability analysis \cite{Sastry1999,Khalil2002}.
In accordance with Eqs.~\eqref{eq:XYZ-F} and \eqref{eq:XYZ-B}, the matrices in the Lyapunov equation are given by
\begin{equation}\label{eq:qfti-x-yF}\textstyle
	\tx(\vk)=-2\mri \tilde{h}(\vk)-\tb_\tr(\vk)\quad \text{and}\quad \ty(\vk)=\tb_\ti(\vk)
\end{equation}
for fermions and by
\begin{equation}\label{eq:qfti-x-yB}\textstyle
	\tx(\vk)=-2\mri\tilde{\tau} \tilde{h}(\vk)+\mri\tilde{\tau} \tb_\ti(\vk)\quad \text{and}\quad
	\ty(\vk)=\tilde{\tau} \tb_\tr(\vk)\tilde{\tau}\quad \text{with}\quad
	\tilde{\tau}=\Psmatrix{0&-\mri\id_{b}\\\mri\id_{b}&0}
\end{equation}
for bosons.

\section{Exponential decay of correlations in quadratic 1D systems}\label{sec:1dQuad}
Let us discuss the proof of Prop.~\ref{prop:1dQuad} in more detail. For a system with finite-range interactions and for sufficiently large $|\vr|$, the right-hand side of Eq.~\eqref{eq:ti_SteadyState} is zero, and, on the left-hand side, the $z_u$ term vanishes as $z_u(\vn,\vec{l})$ is zero for sufficiently large $|\vn|$ or $|\vec{l}|$. For the 1D systems addressed by Prop.~\ref{prop:1dQuad}, without loss of generality, let us consider the decay of $\gamma(r)$ in the positive $x$ direction, i.e., $r>0$.
The $y$ and $z_u$ terms in Eq.~\eqref{eq:ti_SteadyState} shall be zero for $r>R_0$, and $x(n)=0$ for $|n|>d$. Hence, for $r+d>R_0$, Eq.~\eqref{eq:ti_SteadyState} simplifies to
\begin{equation*}
	 \sum_{|n|\leq d}\left[x(-n)\gamma(r+d+n)+\gamma(r+d+n) x^T(n)\right]
	=\sum_{n=0}^{2d}\left[x(d-n)\gamma(r+n)+\gamma(r+n) x^T(n-d)\right]
	= 0.
\end{equation*}
Arranging the elements of the $2b\times 2b$ matrices $\gamma(r)$ in vectors $\vgamma(r)\in \RR^{4b^2}$, we arrive the following \emph{matrix difference equation} for the steady-state covariance matrix,
\begin{equation}\label{eq:1dMatrixDiffAppx}\textstyle
	\sum_{n=0}^R C_n \vgamma(r+n)= \vec{0}\quad \text{with}\quad
	C_n:=x(d-n)\otimes\id_{2b}+\id_{2b}\otimes x(n-d)\quad \text{and}\quad
	R:=2d.
\end{equation}

If $C_R$ is invertible, Eq.~\eqref{eq:1dMatrixDiffAppx} can be solved for $\vgamma(r+R)$ as a linear function of $\vgamma(r),\dotsc,\vgamma(r+R-1)$. As argued in Sec.~\ref{sec:quadratic1d}, the finite dimension of the corresponding transfer matrix \eqref{eq:1dTransferMatrix} implies that, at large $r$, the correlations $\gamma(r)$ can only decay exponentially, converge to a constant or oscillate with a constant amplitude. These three options correspond to transfer-matrix eigenvalues $\beta\in \CC$ with $|\beta| < 1$, $\beta = 1$, and $\beta=e^{\mri \varphi\neq 0}$, respectively. While the transfer matrix may have eigenvalues $|\beta|>1$, the corresponding indefinite exponential increase of correlations can be excluded for physical systems.

This result can also be established when $C_R$ is not invertible by adapting the generating-function method for the analysis of scalar recurrence relations \cite{Wilf2006} to the matrix-difference problem \eqref{eq:1dMatrixDiffAppx}:
The \emph{generating function}
\begin{equation}\label{eq:genFctDef}\textstyle
	\vg(z):=\sum_{r=0}^\infty z^r \vgamma(r)
\end{equation}
exists for $|z|<1$ as, for physical systems, $\vgamma(r)$ is bounded. Multiplication of the generating function with the \emph{matrix pencil} $L(z):=\sum_{n=0}^R z^{R-n} C_n$ yields
\begin{equation}
	L(z)\vg(z)=\sum_{r=0}^\infty z^{r+R}\left[C_0\vgamma(r)+\dotsb+C_R\vgamma(r+R)\right] + \vg_0(z) = \vg_0(z),
\end{equation}
where we have used the matrix difference equation \eqref{eq:1dMatrixDiffAppx}, and we have defined $\vg_0(z):=\sum_{n=1}^R\sum_{r=0}^{n-1}z^{R-n+r}C_n\vgamma(r)$.

The matrix pencil $L(z)$ is called \emph{regular} if there exists at least one point $z$ with $\det L(z)\neq 0$. In Prop.~\ref{prop:1dQuad}, we assume that the steady state is unique. Hence, $L(z)$ must be regular, and, as its degree is $R$, $L(z)$ is invertible except for a finite number of isolated points. In particular, $\vg(0)=\vgamma(0)$ is finite such that we can Taylor expand the generating function
\begin{equation}\label{eq:genFctSol}
	\vg(z)=L^{-1}(z)\vg_0(z)
\end{equation}
at $z=0$ to obtain, according to Eq.~\eqref{eq:genFctDef}, the solution $\vgamma(r)$ from the expansion coefficients. The inverse of $L(z)$ can be expressed in terms of its adjugate divided by its determinant. So $\vg(z)$ is a rational function. The asymptotic behavior of $\vgamma(r)$ is determined by the poles of $\vg(z)$. The \emph{partial fraction decompositions} of any component of the vector $\vg(z)$ yields an expression of the form
\begin{equation}
	g_j(z)=p(z)+\sum_i \frac{f_i(z)}{(1-\alpha_i z)^{n_i}}
\end{equation}
with a finite sum, where $p(z)$ and $f_i(z)$ are polynomials of finite degree, and the degree of $f_i(z)$ is smaller than $n_i$. As explained above, $\vg(z)$ exists for all $|z|<1$, i.e., poles necessarily occur at $|z|\geq 1$ which implies that $|\alpha_i|\leq 1$.
According to the binomial theorem for negative powers and $|\alpha_i z|<1$, the Taylor expansion of the denominators at $z=0$ reads
\begin{equation}\label{eq:genFctTaylor}
	\frac{1}{(1-\alpha_i z)^{n_i}}=\sum_{r=0}^\infty z^r \alpha_i^r \binom{r+n_i-1}{r}.
\end{equation}
For large distances $r$, the binomial coefficient scales like $r^{n_i-1}$. So, a term like Eq.~\eqref{eq:genFctTaylor} in the generating function yields a contribution $\sim r^{n_i-1}\alpha_i^r$ to the steady-state covariance matrix $\gamma(r)$. Multiplying Eq.~\eqref{eq:genFctTaylor} by a term $~z^m$ from the nominator-polynomial $f_i(z)$, corresponds to an inessential shift in $r$. Asymptotically,
\begin{equation}\textstyle
	\gamma(r) \sim r^\nu \alpha_{i_0}^r\quad\text{with $i_0$ such that}\quad |\alpha_{i_0}|=\max_i |\alpha_i|
\end{equation}
and a finite integer power $\nu\geq 0$. As $|\alpha_{i}|\leq 1$, this concludes the proof of Prop.~\ref{prop:1dQuad}.

\section{Critical quasifree bosonic models}\label{sec:criticalBosons}
The statement of Prop.~\ref{prop:quasiFreeB} about the existence of quasifree bosonic models that are critical for $D\geq 2$ dimensions can be established by analyzing a specific purely dissipative model.

\subsection{The model and solution of the Lyapunov equation}\label{sec:criticalBosons-model}
The considered model has one mode per site $\vec{j}$ of the (infinite) $D$-dimensional square lattice $\ZZ^D$. It is characterized by the $4D+1$ linear Lindblad operators
\begin{equation}\label{eq:criticalBosons-L}
	\hL^{(1)}_\vec{j}=\sqrt{2D\eta}\,(\hw_{\vec{j}+}-\mri \hw_{\vec{j}-}),\quad
	\hL^{(2\pm)}_{\vec{j},a}=\hw_{\vec{j}+}+\mri \hw_{(\vec{j}\pm\vec{e}_a)-},\ \ \text{and}\ \
	\hL^{(3\pm)}_{\vec{j},a}=\hw_{\vec{j}+}\pm\hw_{(\vec{j}\pm\vec{e}_a)-}
\end{equation}
per site, where $\vec{e}_a$ are unit vectors for each spatial direction with $a=1,\dotsc,D$.
According to Eqs.~\eqref{eq:H-L-M} and \eqref{eq:ti_L-M}, these Lindblad operators are characterized by the coefficient vectors
\begin{equation}\label{eq:criticalBosons-l}
\begin{aligned}
	&\vec{\ell}^{(1)}(\vr)=\sqrt{2D\eta}\,\delta_{\vr,\vec{0}}\Pmatrix{1\\-\mri},\quad
	\vec{\ell}_a^{(2\pm)}(\vr)=\delta_{\vr,\vec{0}}\Pmatrix{1\\0}+\delta_{\vr,\pm\vec{e}_a}\Pmatrix{0\\\mri},\quad\text{and}\\
	&\vec{\ell}_a^{(3\pm)}(\vr)=\delta_{\vr,\vec{0}}\Pmatrix{1\\0}\pm\delta_{\vr,\pm\vec{e}_a}\Pmatrix{0\\1}.
\end{aligned}
\end{equation}
The Fourier transforms \eqref{eq:qfti_FourierAppx} are
\begin{equation}\label{eq:criticalBosons-lk}
	\tell^{(1)}(\vk)=\sqrt{2D\eta}\,\Pmatrix{1\\-\mri},\quad
	\tell_a^{(2\pm)}(\vk)=\Pmatrix{1\\\mri e^{\mp\mri k_a}},\quad\text{and}\quad
	\tell_a^{(3\pm)}(\vk)=\Pmatrix{1\\\pm e^{\mp\mri k_a}}
\end{equation}
such that the matrix \eqref{eq:qfti_bk} for this model reads
\begin{equation}
	\tb(\vk) = 2D\Pmatrix{\eta + 2& \mri(\eta-c_\vk+s_\vk)\\ -\mri(\eta-c_\vk+s_\vk) & \eta+2},
\end{equation}
where $c_\vk=\sum_a\cos k_a/D$ and $s_\vk=\sum_a\sin k_a/D$.
With the Pauli matrix $\sigma_y=\Psmatrix{0&-i\\i&0}$ as well as $\tb_\tr(\vk)$ and $\tb_\ti(\vk)$ from Eq.~\eqref{eq:qfti_bk}, the matrices $\tx(\vk)$ and $\ty(\vk)$ in the Lyapunov equation \eqref{eq:qftiSteadyState} for this purely dissipative quasifree model are then
\begin{equation}\label{eq:criticalBosons-xy}
	\tx(\vk)\stackrel{\eqref{eq:qfti-x-yB}}{=}\mri \sigma_y\tb_\ti(\vk) = 2D\Pmatrix{c_\vk-\eta& \\ & c_\vk-\eta}\ \ \text{and}\ \ 
	\ty(\vk)\stackrel{\eqref{eq:qfti-x-yB}}{=}\sigma_y\tb_\tr(\vk)\sigma_y = 2D\Pmatrix{\eta+2& \mri s_\vk\\ -\mri s_\vk & \eta+2}.
\end{equation}

As discussed in Ref.~\cite{Barthel2021_12}, the \emph{dissipative gap} $\Delta$ of a quasifree system is determined by the largest eigenvalue real part of the matrix $X$. The eigenvalues of the latter are given by the union of the $\tx(\vk)$ eigenvalues $\forall \vk$. With the doubly degenerate $\tx(\vk)$ eigenvalue $\xi(\vk)=2D\,(c_\vk-\eta)$, we have $\Delta=-\max_\vk\Re\xi(\vk)=D(\eta-1)$. So, the model is stable for $\eta\geq 1$ and the gap closes for $\eta=1$. As $\tx(\vk)$ is proportional to the identity, one can easily solve the Lyapunov equation \eqref{eq:qftiSteadyState} to obtain the solution \eqref{eq:criticalBosons-gk} for $\tgamma(\vk)$.
An inverse Fourier transformation yields the real-space covariance matrix. In the thermodynamic limit, its diagonal and off-diagonal elements read
\begin{subequations}\label{eq:criticalBosons-gr}
\begin{alignat}{5}\label{eq:criticalBosons-grd}
	\gamma_{+,+}(\vr)&=\frac{D}{2(2\pi)^D}\int_{0}^{2\pi}\!\!\! \ud^Dk\, e^{\mri \vk\cdot\vr} \frac{\eta+2}{D\eta-\sum_a \cos k_a}&&=\gamma_{-,-}(\vr)\quad\text{and}\\
	\label{eq:criticalBosons-gro}
	\gamma_{+,-}(\vr)&=\frac{\mri}{2(2\pi)^D}\int_{0}^{2\pi}\!\!\! \ud^Dk\,e^{\mri \vk\cdot\vr} \frac{\sum_a \sin k_a}{D\eta-\sum_a \cos k_a}&&=-\gamma_{-,+}(\vr).
\end{alignat}
\end{subequations}

\subsection{Exponential decay in \texorpdfstring{$D=1$}{D=1} dimensions}
For $D=1$ dimensions, the Fourier integrals \eqref{eq:criticalBosons-gr} can be evaluated exactly by defining the complex variable $z:=e^{\mri k}$ and using \emph{Cauchy's  residue theorem}. For $\eta>1$, the diagonal matrix element is 
\begin{equation}\label{eq:criticalBosons-grD1d}
	\gamma_{+,+}(r)=\frac{1}{4\pi}\int_{0}^{2\pi}\!\!\! \ud k\, e^{\mri kr} \frac{\eta+2}{\eta-\cos k}
	= -\frac{\eta+2}{2\pi\mri} \oint_{|z|=1} \!\!\!\ud z\, \frac{z^r}{z^2-2\eta z+1}
	= (\eta+2)\frac{z_-^r}{2\sqrt{\eta^2-1}}.
\end{equation}
With $z_\pm:=\eta\pm\sqrt{\eta^2-1}$ and $z^2-2\eta z+1=(z-z_+)(z-z_-)$, it has been used in the last step that the pole at $z=z_-$ is inside the unit circle, while $z_+$ is outside, and that the residue of the integrand at $z=z_-$ is $z_-^r/(z_--z_+)=-z_-^r/2\sqrt{\eta^2-1}$.
Similarly, the off-diagonal element is
\begin{equation}\label{eq:criticalBosons-grD1o}
	\gamma_{+,-}(r)=\frac{\mri}{4\pi}\int_{0}^{2\pi}\!\!\! \ud k\, e^{\mri kr} \frac{\sin k}{\eta-\cos k}
	= \frac{\mri}{4\pi} \oint_{|z|=1} \!\!\!\ud z\, z^{r-1} \frac{z^2-1}{z^2-2\eta z+1}
	= -\frac{1}{2}\,z_-^{r-1}\frac{1-z_-^2}{2\sqrt{\eta^2-1}}.
\end{equation}
In this case, the residue of the integrand at $z=z_-$ is $z_-^{r-1}(1-z_-^2)/2\sqrt{\eta^2-1}$.

For $\eta>1$ the system is gapped and $|z_-|<1$. Thus, Eqs.~\eqref{eq:criticalBosons-grD1d} and \eqref{eq:criticalBosons-grD1o} imply an exponential decay in $r$, in agreement with Props.~\ref{prop:1dQuad} and \ref{prop:gapped}. Furthermore, $\lim_{\eta\to 1}z_-=1$ such that the correlation length diverges as the gap closes. Additionally, the amplitude of $\gamma_{+,+}(r)$ will diverge (the model requires infra-red regularization), but $\gamma_{+,-}(r)$ converges to the constant $\lim_{\eta\to 1}\gamma_{+,-}(r)=-1/2$. In any case, no power-law decay is possible, as predicted by Prop.~\ref{prop:1dQuad}.
\begin{figure*}[t]
	\begin{tabular}{c@{\hskip 0.05\textwidth}c@{\hskip 0.05\textwidth}c}
	\includegraphics[width=0.27\textwidth]{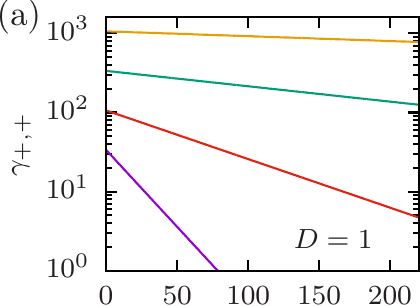}&
	\includegraphics[width=0.27\textwidth]{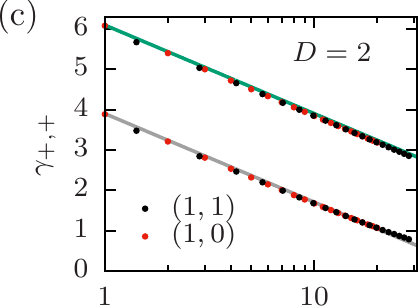}&
	\includegraphics[width=0.27\textwidth]{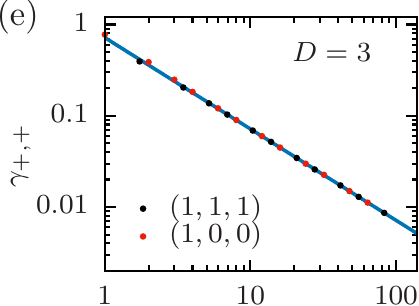}\vspace{1.2em}\\
	\includegraphics[width=0.27\textwidth]{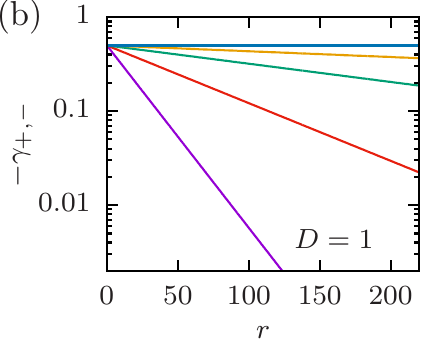}&
	\includegraphics[width=0.27\textwidth]{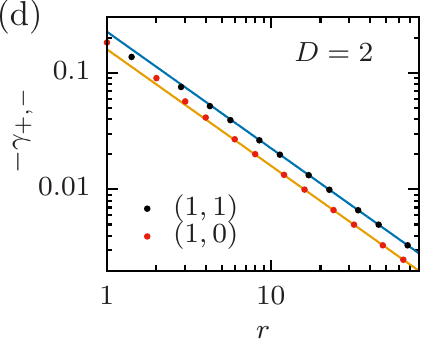}&
	\includegraphics[width=0.27\textwidth]{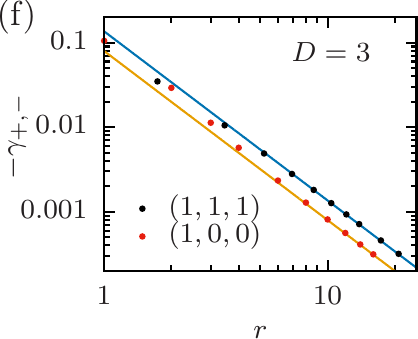}
	\end{tabular}
	\caption{\label{fig:criticalBosons}\textbf{Decay of correlations for gapless bosons.} For the purely dissipative model described in Appx.~\ref{sec:criticalBosons-model}, the plots show the decay of diagonal and off-diagonal covariance-matrix elements \eqref{eq:criticalBosons-gr} as a function of the distance $r$. The dissipative gap closes for $\eta\to 1$.
	(a) Exponential decay \eqref{eq:criticalBosons-grD1d} of $\gamma_{+,+}$  for $D=1$ dimensions and $\eta-1=10^{-3},10^{-4},10^{-5},10^{-6}$.
	(b) Exponential decay \eqref{eq:criticalBosons-grD1o} of $\gamma_{+,-}$  for $D=1$ dimensions and $\eta-1=10^{-3},10^{-4},10^{-5},10^{-6},0$.
	(c) Logarithmic decay \eqref{eq:criticalBosons-grD2d} of $\gamma_{+,+}$  for $D=2$ dimensions and $\eta-1=10^{-4},10^{-6}$. Dots indicate exact values \eqref{eq:criticalBosons-gr} for the $x$ direction ``$(1,0)$'' and the diagonal direction ``$(1,1)$''.
	(d) $1/r$ decay \eqref{eq:criticalBosons-grD2o} of $\gamma_{+,-}$  for $D=2$ dimensions and $\eta=1$.
	(e) $1/r$ decay \eqref{eq:criticalBosons-grD3d} of $\gamma_{+,+}$  for $D=3$ dimensions and $\eta=1$. Dots indicate exact values \eqref{eq:criticalBosons-gr} for the $x$ direction ``$(1,0,0)$'' and the diagonal direction ``$(1,1,1)$''.
	(d) $1/r^2$ decay \eqref{eq:criticalBosons-grD3o} of $\gamma_{+,-}$  for $D=3$ dimensions and $\eta=1$.
	}
\end{figure*}

\subsection{Logarithmic and power-law decays in \texorpdfstring{$D=2$}{D=2} dimensions}
To study the decay of correlations in two dimensions, we can employ polar coordinates
\begin{equation}
	\vk^T=(k_1,k_2)=(k\cos\varphi_k,k\sin\varphi_k)\quad\text{and}\quad
	\vr^T=(r\cos\varphi_r,r\sin\varphi_r).
\end{equation}
For the Fourier integral \eqref{eq:criticalBosons-gr}, we can first expand $\tgamma(\vk)$ in a \emph{multipole series}
\begin{equation}\label{eq:criticalBosons-gkD2multipole}\textstyle
	\tgamma(\vk)=\sum_{m-\infty}^\infty\tgamma^{(m)}(k)\,e^{\mri m\varphi_k}
\end{equation}
such that the Fourier transform reduces to the sum of \emph{Hankel transforms}
\begin{equation}\label{eq:criticalBosons-grD2Hankel}
	\gamma(\vr)=\frac{1}{(2\pi)^2}\int\ud^2k\,e^{\mri \vk\cdot\vr} \tgamma(\vk)
	=\frac{1}{2\pi}\sum_{m-\infty}^\infty \mri^m e^{\mri m\varphi_r} \int\ud k\,  k\,\tgamma^{(m)}(k)\,J_m(kr),
\end{equation}
where $J_m(q)$ denotes the \emph{Bessel function of the first kind} that arises from the integral over $\varphi_k$,
\begin{equation*}\textstyle
	\frac{1}{2\pi}\int_0^{2\pi}\ud\varphi_k\,e^{\mri q\cos(\varphi_k-\varphi_r)+ \mri m\varphi_k}
	=\frac{1}{2\pi}\,e^{\mri m\varphi_r}\int_0^{2\pi}\ud\psi\,e^{\mri q\cos\psi+ \mri m\psi}
	=\mri^m e^{\mri m\varphi_r} J_m(q).
\end{equation*}

For the large-$r$ asymptotics of the diagonal matrix elements, the leading contribution is the $m=0$ component, and we can approximate
\begin{equation}\label{eq:criticalBosons-gkD2d}
	\tgamma_{+,+}(\vk)= \frac{\eta+2}{2\eta-\cos k_1-\cos k_2} \approx 2 \Theta(K-k)\, \frac{\eta+2}{\veps+k^2},
\end{equation}
such that we just have the component $\tgamma^{(0)}_{+,+}$. We account for the Brillouin zone boundary at $|k_a|=\pi$ by the momentum cutoff $K\sim\pi$ in the Heaviside step function. $\veps:=4(\eta-1)$ is proportional to the dissipative gap. Plugging the approximation \eqref{eq:criticalBosons-gkD2d} into Eq.~\eqref{eq:criticalBosons-grD2Hankel}, we obtain
\begin{align}\nonumber
	\gamma_{+,+}(\vr)
	&\textstyle \approx\frac{\eta+2}{\pi} \int_0^K\ud k\,  \frac{k}{\veps+k^2}\,J_0(kr)
	 =\frac{\eta+2}{\pi} \int_0^{Kr}\ud q\,  \frac{q}{\veps r^2 +q^2}\,J_0(q)\\ \label{eq:criticalBosons-grD2d-split}
	&\textstyle \approx \frac{\eta+2}{\pi} \left[\int_0^Q \ud q\,  \frac{q}{\veps r^2 +q^2}
	   + \sqrt{\frac{2}{\pi}}\int_Q^{Kr} \ud q\, \frac{1}{q^{3/2}}\,\cos\big(q-\frac{\pi}{4}\big) \right].
\end{align}
The integral has been split into two parts, and we can use $Q=1/2$. For small $q:=kr$ (and large $r$) we use that $J_0(q)=1-q^2/4+\mc{O}(q^4)$. This first part gives $\frac{1}{2}\ln\left(1+Q^2/(\veps r^2)\right)\approx -\frac{1}{2}\ln\veps-\ln(2r)$. The nonzero gap ($\veps$) regularizes this part of the integral; similarly, one could regularize by introducing an infra-red momentum cutoff. For the large-$q$ part in Eq.~\eqref{eq:criticalBosons-grD2d-split}, we have used that $J_0(q)\approx \sqrt{2/\pi q}\,\cos(q-\pi/4)$ for $q\gg 1/4$. Its absolute value is bounded by $\int_Q^{Kr} \ud q\, q^{-3/2}=2/\sqrt{Q}-2/\sqrt{Kr}$, i.e., a subleading power-law decay in $r$. Sending $r$ to infinity for this part, the integral can be evaluated by employing the \emph{Fresnel integrals} $S(x)$ and $C(x)$, giving $2\sin(1/2)+ 2\cos(1/2) +2\sqrt{\pi}\left[S(\pi^{-1/2})-C(\pi^{-1/2})\right]\approx 1.0909$. In combination,
\begin{equation}\label{eq:criticalBosons-grD2d}
	\tgamma_{+,+}(\vr)\approx \frac{\eta+2}{\pi} \left( -\ln\sqrt{\eta-1}-0.5159-\ln r\right),
\end{equation}
i.e., a logarithmic decay in $r$ starting from a constant that diverges when closing the gap ($\eta\to 1$).

The off-diagonal element $\tgamma_{+,-}(\vk)$ in Eq.~\eqref{eq:criticalBosons-gk} is anti-symmetric with respect to reflection $\vk\leftrightarrow-\vk$ such that the $m=0$ component in the multipole expansion \eqref{eq:criticalBosons-gkD2multipole} is zero and $m=\pm 1$ gives the leading contribution to the large-$r$ asymptotics of $\gamma_{+,-}(\vr)$. The corresponding Hankel transform does not require regularization so that we can set $\eta=1$ from the beginning, i.e., we approximate
\begin{equation}\label{eq:criticalBosons-gkD2o}
	\tgamma_{+,-}(\vk)= \frac{\mri}{2}\,\frac{\sin k_1+\sin k_2}{2-\cos k_1-\cos k_2} \approx \Theta(K-k)\, \frac{\sqrt{2}\,\mri\,\sin(\varphi_k+\pi/4)}{k}.
\end{equation}
This implies $\tgamma^{(0)}_{+,-}=0$, $\tgamma^{(1)}_{+,-}=\Theta(K-k)\, \sqrt{2}\, e^{\mri \pi/4}/k$, and $\tgamma^{(-1)}_{+,-}=-\tgamma^{(1)*}_{+,-}$. Plugging the approximation \eqref{eq:criticalBosons-gkD2o} into Eq.~\eqref{eq:criticalBosons-grD2Hankel}, we obtain
\begin{equation}\label{eq:criticalBosons-grD2o}
	\gamma_{+,-}(\vr)
	 \approx - \frac{\sin(\varphi_r+\pi/4)}{\sqrt{2}\,\pi}\int_0^K\ud k\, J_1(kr)
	 = - \frac{\sin(\varphi_r+\pi/4)}{\sqrt{2}\,\pi\, r}[1-J_1(Kr)].
\end{equation}
As $J_1(Kr)\to 0$ for large $r$, the off-diagonal elements follow a $1/r$ decay, i.e., the model is critical. Figure~\ref{fig:criticalBosons} compares the asymptotic analytical results to numerical data.

\subsection{Power-law decays in \texorpdfstring{$D=3$}{D=3} dimensions}
To study the decay of correlations in three dimensions, we can employ spherical coordinates
\begin{align*}
	\vk^T&=(k_1,k_2,k_3)=(k\sin\theta_k\cos\varphi_k,k\sin\theta_k\sin\varphi_k,k\cos\theta_k)\quad\text{and}\\
	\vr^T&=(r\sin\theta_r\cos\varphi_r,r\sin\theta_r\sin\varphi_r,r\cos\theta_r).
\end{align*}
For the Fourier integral \eqref{eq:criticalBosons-gr}, we can expand $\tgamma(\vk)$ in a \emph{multipole series} over spherical harmonics
\begin{equation}\label{eq:criticalBosons-gkD3multipole}\textstyle
	\tgamma(\vk)=\sum_{\ell=0}^\infty\sum_{m=-\ell}^\ell\tgamma^{(\ell,m)}(k)\,Y_{\ell,m}(\theta_k,\varphi_k)
\end{equation}
such that the Fourier transform reduces to the sum of \emph{Hankel transforms}
\begin{equation}\label{eq:criticalBosons-grD3Hankel}
	\gamma(\vr)=\frac{1}{(2\pi)^3}\int\ud^2k\,e^{\mri \vk\cdot\vr} \tgamma(\vk)
	=\frac{1}{(2\pi)^{3/2}} \sum_{\ell,m} \mri^\ell
	    \frac{Y_{\ell,m}(\theta_r,\varphi_r)}{\sqrt{r}}\int\ud k\,  k^{3/2}\,\tgamma^{(\ell,m)}(k)\,J_{\ell+1/2}(kr),
\end{equation}
where $J_m(q)$ denotes the \emph{Bessel function of the first kind} arising from the integral over $\varphi_k$ and $\theta_k$:
\begin{gather*}\textstyle
	e^{\mri \vk\cdot\vr} = 4\pi\sum_{\ell,m}\mri^\ell\sqrt{\frac{\pi}{2kr}}\,J_{\ell+1/2}(kr)
	                          Y_{\ell,m}(\theta_r,\varphi_r)Y^*_{\ell,m}(\theta_k,\varphi_k)\quad\text{and}\\\textstyle
	\int_0^\pi\ud\theta \sin\theta\int_0^{2\pi}\ud\varphi\, Y_{\ell,m}(\theta,\varphi)Y^*_{\ell',m'}(\theta,\varphi)
	  =\delta_{\ell,\ell'}\delta_{m,m'}\\\textstyle
	\Rightarrow\quad
	\int_0^\pi\ud\theta_k\sin\theta_k\int_0^{2\pi}\ud\varphi_k\,e^{\mri \vk\cdot\vr} Y_{\ell,m}(\theta_k,\varphi_k)
	 = (2\pi)^{3/2} \,\mri^\ell\,J_{\ell+1/2}(kr)/\sqrt{kr}.
\end{gather*}

We only need to consider the gapless system $\eta=1$ as the corresponding Fourier integrals \eqref{eq:criticalBosons-gr} do not require regularization.
For the large-$r$ asymptotics of the diagonal matrix elements, the leading contribution is the $\ell=0$ component, and we can approximate
\begin{equation}\label{eq:criticalBosons-gkD3d}
	\tgamma_{+,+}(\vk)= \frac{9}{2}\,\frac{1}{3-\cos k_1-\cos k_2-\cos k_3} \approx \Theta(K-k)\, \frac{9}{k^2},
\end{equation}
so that we just have the component $\tgamma^{(0,0)}_{+,+}$. The (inessential) momentum cutoff $K\sim\pi$ represents the Brillouin zone boundary. Plugging the approximation \eqref{eq:criticalBosons-gkD3d} into Eq.~\eqref{eq:criticalBosons-grD3Hankel}, we obtain
\begin{equation}\label{eq:criticalBosons-grD3d}
	\gamma_{+,+}(\vr)
	 \approx  \frac{9}{(2\pi)^{3/2}} \frac{1}{\sqrt{r}}\int_0^K\ud k\,  J_{1/2}(kr)/\sqrt{k}
	 =\frac{9}{2\pi^2}\, \frac{1}{r}\int_0^{K}\ud k\,  \sin(kr)/k = \frac{9}{2\pi^2}\, \frac{1}{r} \operatorname{Si}(Kr).
\end{equation}
The so-called sine integral $\operatorname{Si}(Kr)\to \pi/2$ for large $r$, implying that the diagonal matrix elements decay as $9/(4\pi r)$ at large $r$.

The off-diagonal element $\tgamma_{+,-}(\vk)$ in Eq.~\eqref{eq:criticalBosons-gk} is anti-symmetric with respect to reflection $\vk\leftrightarrow-\vk$. Hence, the $\ell=0$ component in the multipole expansion \eqref{eq:criticalBosons-gkD3multipole} is zero and the $\ell=1$ components give the leading contribution to the large-$r$ asymptotics of $\gamma_{+,-}(\vr)$. We can approximate
\begin{equation}\label{eq:criticalBosons-gkD3o}
	\tgamma_{+,-}(\vk)= -\frac{\mri}{2}\,\frac{\sin k_1+\sin k_2+\sin k_3}{3-\cos k_1-\cos k_2-\cos k_3}
	\approx \mri\,\Theta(K-k)\, \frac{\sin k_1+\sin k_2+\sin k_3}{k^2},
\end{equation}
such that
\begin{equation}\textstyle
	\tgamma^{(0,0)}_{+,-}=0,\quad
	\tgamma^{(1,0)}_{+,-}=2\mri\,\Theta(K-k)\sqrt{\frac{\pi}{3}}\,\frac{1}{k},\quad\text{and}\quad
	\tgamma^{(1,\pm 1)}_{+,-}=-\Theta(K-k)\sqrt{\frac{2\pi}{3}}\,(1\pm \mri)\,\frac{1}{k}.
\end{equation}
Plugging the approximation \eqref{eq:criticalBosons-gkD3o} into Eq.~\eqref{eq:criticalBosons-grD3Hankel}, the $r$-dependent part of the $\ell=1$ component is
\begin{equation}\textstyle
	\frac{1}{\sqrt{r}}\,\int_0^K\ud k\, \sqrt{k}\,J_{3/2}(kr)
	=\frac{1}{r^2}\,\int_0^{Kr}\ud q\, \sqrt{q}\,J_{3/2}(q)
	=\sqrt{\frac{2}{\pi}}\left[\operatorname{Si}(Kr)-\sin(Kr)\right]\, \frac{1}{r^2}.
\end{equation}
For large $r$, the sine integral converges to $\pi/2$, i.e., the term in the bracket oscillates in $[\frac{\pi}{2}-1,\frac{\pi}{2}+1]$. In conclusion, the off-diagonal elements follow the $1/r^2$ decay
\begin{equation}\label{eq:criticalBosons-grD3o}
	\gamma_{+,-}(\vr)\approx \frac{-1}{\pi^{3/2}\sqrt{3}}
	  \left[Y_{(1,0)}(\theta_r,\phi_r)+\frac{\mri-1}{\sqrt{2}}\,Y_{(1,1)}(\theta_r,\phi_r)+\frac{\mri+1}{\sqrt{2}}\,Y_{(1,-1)}(\theta_r,\phi_r)\right]
	  \left[\frac{\pi}{2}-\sin(Kr)\right]\, \frac{1}{r^2}.
\end{equation}
The remaining $K$ dependence indicates that this integral is somewhat sensitive to how we handle details at the Brillouin zone boundary. However, the $1/r^2$ decay, the coefficient, and the angular dependence are confirmed by the comparison to numerical data in Fig.~\ref{fig:criticalBosons}.

\section{An illustrative quadratic fermion model}\label{sec:XYchain}
Using the Jordan-Wigner transformation \cite{Jordan1928,Lieb1961}, the Hamiltonian of the spin-1/2 XY chain is mapped to the fermionic form
\begin{align}\nonumber
	\hH&
	  =\sum_{j}\left(\ha_j^\dag\ha_{j+1}^\pdag+\alpha\ha_j^\dag\ha_{j+1}^\dag +H.c.\right)
	   -\mu\sum_j\left(\ha_j^\dag\ha_j^\pdag-1/2\right)\\\label{eq:XYchain2}
	 &= \sum_j\left[\hvw^T_j\left(\sigma_y+\mri\alpha\sigma_x\right)\hvw_{j+1}
	   -\frac{\mu}{2}\hvw^T_j\sigma_y\hvw_j\right].
\end{align}
Here and in the following, $\sigma_x$, $\sigma_y$, and $\sigma_z$ denote the Pauli matrices and $\hvw^T_j:=(\hw_{j+},\hw_{j-})$ is the vector of fermionic Majorana operators for site $j$. Hence, the Hamiltonian coupling coefficients $h(r)$ for this translation-invariant system are [cf.~\eqref{eq:ti_H-G}]
\begin{equation}\label{eq:XYchain-h}\textstyle
	h(r)=\frac{1}{2}
	\begin{cases}
	-\mu\sigma_y                   &\text{for}\ r=0,\\
	\sigma_y\mp\mri\alpha\sigma_x  &\text{for}\ r=\pm1\\
	\end{cases}\,\ \text{such that}\ \
	\tilde{h}(k)\stackrel{\eqref{eq:qfti_FourierAppx}}{=}\left(\cos k -\frac{\mu}{2}\right)\,\sigma_y-\alpha\sin k\, \sigma_x.
\end{equation}

Let us first discuss the quasifree case with linear Lindblad operators
\begin{equation}
	\hL_j = \sqrt{\eta/2}\,\left[\ha_{j}+\ha^\dag_{j}+e^{\mri \phi}\left(\ha_{j+1}+\ha^\dag_{j+1}\right)\right]
	 =\sqrt{\eta}\,\left(\hw_{j+}+e^{\mri \phi}\hw_{(j+1)+}\right)
\end{equation}
as introduced in Ref.~\cite{Eisert2010_12}. According to Eq.~\eqref{eq:ti_L-M}, the corresponding coefficient vector is
\begin{align}\label{eq:XYchain-l}
	\vec{\ell}(r)&\textstyle
	 =\sqrt{\eta}\,\left(\delta_{r,0}+\delta_{r,1} e^{\mri\phi}\right)\,\Pmatrix{1\\0}
	\,\ \text{such that}\ \
	\tell(k)\stackrel{\eqref{eq:qfti_FourierAppx}}{=} \sqrt{\eta}\,\left(1+ e^{\mri(\phi-k)}\right)\,\Pmatrix{1\\0}
	\quad\text{and}\\
	\tb(\vk)&\textstyle
	 \stackrel{\eqref{eq:qfti_bk}}{=} \eta\left[1+\cos(\phi-k)\right]\,(\sigma_z+\id).
\end{align}
In this way, we have determined the matrices
\begin{alignat*}{4}
	\tx(\vk)&\stackrel{\eqref{eq:qfti-x-yF}}{=}-2\mri\tilde{h}(k)-\tb_\tr(k)
	  &&= \mri\,(\mu-2\cos k)\sigma_y +2\mri\alpha\sin k\, \sigma_x - \eta(1+\cos\phi\cos k)\,(\sigma_z+\id),\\
	\ty(\vk)&\stackrel{\eqref{eq:qfti-x-yF}}{=}\tb_\ti(k)
	  &&= -\mri \eta \sin\phi\sin k\,(\sigma_z+\id)
\end{alignat*}
in the Lyapunov equation \eqref{eq:qftiSteadyState}.

The $\tx(\vk)$ eigenvalues
\begin{equation}\label{eq:XYchain-qf-xi}
	\xi_\pm(k) = \frac{1}{2}\left(f(k)\pm\sqrt{f^2(k)-|2c(k)|^2}\right)
\end{equation}
with $f(k):=-2\eta(1+\cos\phi\cos k)$ and $c(k):=\mu-2\cos k+2\mri\alpha\sin k$
determine the dissipative gap $\Delta$ of the Liouvillian \cite{Barthel2021_12}. Here, $\Delta=-\max_k\Re\xi_+(k)$. Let us assume that the dissipation is on ($\eta>0$), and that the Hamiltonian is gapped, where the latter implies $c(k)\neq 0$ $\forall k$. Then, the dissipative gap $\Delta$ closes for $\phi=0$ and $\phi=\pi$ at quasimomenta $k=\pi$ and $k=0$, respectively. So, this is a nice example, where dissipation closes the gap. 

As the eigenvalues $\xi_\pm(k)$ are nonzero except for isolated points, the Lyapunov equation \eqref{eq:qftiSteadyState} has a unique solution, determined by inversion of $\tx(k)\otimes\id_2+\id_2\otimes \tx(-k)$. It is given by
\begin{equation}
	\tgamma(k)=\frac{\id}{2\mri}\,\frac{\sin\phi\sin k}{1+\cos\phi\cos k}.
\end{equation}
Interestingly, $\tgamma(k)$ is independent of the Hamiltonian parameters \cite{Eisert2010_12}. In the thermodynamic limit, the real-space covariance matrix is obtained by a Fourier integral which can be evaluated exactly by defining the complex variable $z:=e^{\mri k}$ and using \emph{Cauchy's  residue theorem}:
\begin{equation}
	\gamma(r)=\frac{\id}{2\pi}\int_0^{2\pi}\ud k\,e^{\mri kr}\tgamma(k)
	=\frac{\sin\phi\, \id}{4\pi\mri} \oint_{|z|=1} \!\!\!\ud z\, z^{r-1}\,\frac{1-z^2}{z^2\cos\phi+2z+\cos\phi}.
\end{equation}
The denominator of the integrand can be written as $(z-z_+)(z-z_-)\cos\phi$ with $z_\pm:=-(1\pm|\sin\phi|)/\cos\phi$. Of these two pole locations, $|z_-|\leq 1$ and $|z_+|\geq 1$, i.e., $z_-$ is inside the unit circle, approaching its boundary for $\phi\to 0,\pi$. The residue of the integrand at $z=z_-$ is $z_-^{r-1}(1-z_-^2)/[(z_--z_+)\cos\phi]$ such that the integral evaluates to
\begin{equation}\label{eq:XYchain-gr}
	\gamma(r)=\frac{\id}{2}\, \frac{\sin\phi}{1+|\sin\phi|}\, z_-^{r-1}.
\end{equation}
This describes an exponential decay of correlations. While the correlation length $-1/\ln|z_-|$ diverges for $\phi\to 0,\pi$, the prefactor ${\sin\phi}/(1+|\sin\phi|)$ goes simultaneously to zero. The model becomes gapless but is never critical. This is in agreement with Prop.~\ref{prop:quasiFreeF}.

Let us now include the additional quadratic and Hermitian Lindblad operators
\begin{equation}
	\hM_j=\sqrt{\zeta}\,\left(2\ha_j^\dag\ha_j^\pdag-1\right)
	=\mri\sqrt{\zeta}\,\left(\hw_{j-}\hw_{j+}-\hw_{j+}\hw_{j-}\right)
	=\sqrt{\zeta}\,\hvw_j^T\sigma_y\hvw_j.
\end{equation}
The corresponding coefficient matrix in Eq.~\eqref{eq:ti_L-M} is $m(r,r')= \delta_{r,0}\delta_{r',0}\sqrt{\zeta}\,\sigma_y$.
With $h(r)$ from Eq.~\eqref{eq:XYchain-h} and
\begin{equation}
	b(r)\stackrel{\eqref{eq:ti_B}}{=}\sum_n \vec{\ell}(r-n)\vec{\ell}^\dag(-n)
	\stackrel{\eqref{eq:XYchain-l}}{=}\eta\,(\sigma_z+\id)
	\begin{cases}
	1&\text{for}\ r=0,\\
	e^{\pm\mri\phi}/2&\text{for}\ r=\pm 1,
	\end{cases}
\end{equation}
the $x$ matrix in the steady-state equation \eqref{eq:ti_SteadyState} reads
\begin{equation}
	x(r)\stackrel{\eqref{eq:XYZ-F}}{=}-2\mri h(r)-b_\tr(r)-2\sum_n m(r,n)m(n,0) = 
	\begin{cases}
	-\Psmatrix{2\zeta+2\eta&-\mu\\\mu&2\zeta} & \text{for}\ r=0,\\
	\Psmatrix{-\eta\cos\phi&-1\mp\alpha\\1\mp\alpha&0}&\text{for}\ r=\pm 1.
	\end{cases}
\end{equation}
According to Prop.~\ref{prop:1dQuad}, the solution $\gamma(r)$ decays asymptotically as $\beta^r$. In the particular model, we have $d=2$ in the matrix difference equation \eqref{eq:1dMatrixDiffAppx} and the matrix $C_2=x(-1)\otimes \id_2+\id_s\otimes x(1)$ with determinant $4\eta^2(\alpha^2-1)\cos^2\phi$ is invertible as long as $\eta>0$, $|\alpha|\neq 1$ and $\phi\neq\pm\pi/2$. In this case, $\beta$ is simply given by an eigenvalue of the transfer matrix
\begin{equation}\label{eq:XYchain-transMat}
	T=\Pmatrix{-C_2^{-1}C_1&-C_2^{-1}C_0\\\id_4&}
\end{equation}
as discussed in Sec.~\ref{sec:quadratic1d}.

The left panel of Fig.~\ref{fig:XYchain} shows the exponential decay of $\gamma(r)$ for $\mu=0$, $\alpha=1/5$, $\eta=1$ and $\phi=2\pi/5$. The curve for the quasifree case ($\zeta=0$) corresponds to the solution in Eq.~\eqref{eq:XYchain-gr} with $\beta=z_-\approx-0.1584$. This is consistent with the spectrum $\{\beta_i\}$ of the transfer matrix \eqref{eq:XYchain-transMat}. Its eigenvalues with $|\beta_i|\leq 1$ are $-0.1584$ and $\pm0.8165\mri$. The curve for the quadratic case with $\zeta=1/4$ is obtained by solving Eq.~\eqref{eq:ti_SteadyState} for a large system size, $\Ns=400$. The observed exponential decay is governed by the eigenvalue $-0.0194\pm0.5634\mri$ of the transfer matrix \eqref{eq:XYchain-transMat}. The transfer matrix has a second eigenvalue $-0.1041$ with modulus smaller than one.

The right panel of Fig.~\ref{fig:XYchain} shows the dissipative gap $\Delta$ as a function of $\phi$ for $\mu=0$, $\alpha=1/2$, $\eta=1$, and $\zeta=0$ as determined by Eq.~\eqref{eq:XYchain-qf-xi}. For these parameters, the maximum of $\Re\xi_+(k)$ is located at $k=\pi$ for $\phi\in[-\pi/2,\pi/2]$ and at $k=0$ for $\phi\in[\pi/2,3\pi/2]$. Hence, the gap is simply $\Delta=\eta\,(1-\cos\phi)$ for $\phi\in[-\pi/2,\pi/2]$ and $\Delta=\eta\,(1+\cos\phi)$ for $\phi\in[\pi/2,3\pi/2]$. So, as already discussed below Eq.~\eqref{eq:XYchain-qf-xi}, the gap closes at $\phi=0$ and $\phi=\pi$. However, according to Prop.~\ref{prop:phaseUnity}, these are not phase transition points. For example, we can connect the points $\phi=\pi/4$ and $\phi=9\pi/4$ with a continuous path of gapped Liouvillians by switching on the additional Lindblad operators $\hL_{j\pm}=\sqrt{\kappa}\,\hw_{j\pm}$ with $\kappa=0\to 1$ to increase the gap by one, then tuning $\phi=\pi/4\to 9\pi/4$, and finally switching off the additional terms with $\kappa=1\to 0$ as indicated in the figure.

\twocolumngrid
\bibliographystyle{prsty.tb.title}

\end{document}